\documentclass[12pt]{article}
\usepackage[english]{babel}
\usepackage[tbtags]{amsmath}
\usepackage{amsfonts,amssymb,mathrsfs,amscd}
\usepackage[affil-it]{authblk}
\usepackage{hyperref}

\begin{document}

\title{Higher order traps for some strongly degenerate quantum control systems}

\author{Boris O. Volkov\footnote{E-mail: \href{mailto:borisvolkov1986@gmail.com}{borisvolkov1986@gmail.com} URL: \href{https://www.mathnet.ru/php/person.phtml?&personid=94935&option_lang=eng}{mathnet.ru/eng/person94935} ORCID \href{https://orcid.org/0000-0002-6430-9125}{0000-0002-6430-9125}}
~and~Alexander N. Pechen\footnote{E-mail: \href{mailto:apechen@gmail.com}{apechen@gmail.com} URL: \href{https://www.mathnet.ru/eng/person17991}{mathnet.ru/eng/person17991} ORCID \href{https://orcid.org/0000-0001-8290-8300}{0000-0001-8290-8300}}\\
Department of Mathematical Methods for Quantum Technologies, Steklov Mathematical Institute of Russian Academy of Sciences, 8 Gubkina str., Moscow, 119991, Russia}
\date{}

\maketitle

\makeatletter
\renewcommand{\@makefnmark}{}
\makeatother
\footnotetext{This work is supported by the Russian Science Foundation under grant \textnumero~22-11-00330,
\url{https://rscf.ru/en/project/22-11-00330/}.}

Control of quantum systems attracts high interest due to fundamental reasons and applications in quantum technologies~\cite{Koch_2022}. Controlled dynamics of an $N$-level closed quantum system is described by Schr\"odinger equation $$i\dot U_t^f=(H_0+f(t)V)U_t^f$$ with the initial condition $U_{t=0}^f=\mathbb I$ for unitary evolution operator $U_t^f$ in the Hilbert space ${\cal H}=\mathbb C^N$. Here $H_0=H_0^\dagger$ and $V=V^\dagger$ are free and interaction Hamiltonians (Hermitian operators in ${\cal H}$), $f\in\mathfrak{H}^0:= L_2([0,T];\mathbb R)$ is a control function, $T>0$ is some target time. Consider Mayer type quantum control objective functional of the form $J_O(f)={\rm Tr}(OU_T^f\rho_0 U_T^{f\dagger})\to\max$, where $\rho_0$ is the initial density matrix (a Hermitian operator in $\cal H$ such that $\rho_0\ge 0$ and ${\rm Tr}\rho_0=1$) and $O=O^\dagger$ is target operator (a Hermitian operator in $\cal H$).  A quantum system $(H_0,V)$ is called {\it completely controllable} if there exist such time $T_{\rm min}$ that for all $T\geq T_{\rm min}$ and $U\in U(N)$ there exists a control $f\in \mathfrak{H}^0$ such that $U=U_T^fe^{i\alpha}$ for some $\alpha\in \mathbb{R}$. 

An important for quantum control question is to establish, for a controlled system, whether the objective has trapping behaviour or no~\cite{Rabitz2004}. An {\it $n$-th order trap} for the objective functional $J_O$, where $n\geq2$, is a control $f_0\in \mathfrak{H}^0$ such that
(a) $f_0$ is not a point of global extremum of $J_O$ and (b) the Taylor expansion of the objective functional at the point $f_0$ has the form
$$
J_O(f_0+\delta f)=J_O(f_0)+\sum\limits_{j=2}^{n} \frac 1{j!}J^{(j)}_O(f_0)(\delta f,\ldots,\delta f)+o(\|\delta f\|^{n})\text{\, as $\|\delta f\|\rightarrow 0$,}
$$
where the non-zero functional $R(\delta f):=\sum\limits_{j=2}^{n} \frac 1{j!}J^{(j)}_O(f_0)(\delta f,\ldots,\delta f)$ is such that for any $\delta f\in \mathfrak{H}^0$ there exists $\varepsilon>0$ such that
$R(t\delta f)\leq 0$ for all $t\in (-\varepsilon,\varepsilon)$. The analysis of traps is important since traps, if they would exist, would determine the level of difficulty for the search for globally optimal controls, including in practical applications. The absence of traps for 2-level quantum systems ($N=2$) was proved in~\cite{PechenPRA2012,PechenRMS2015,VolkovJPA2021}. In~\cite{Pechen2011,Pechen2012}, some examples of {\it third order traps} were constructed for special $N$--level degenerate quantum systems with $N\ge 3$. Traps were discovered for some systems with $N\geq 4$ in~\cite{deFouquieresSchirmer}. In this work, we prove the existence of traps of an arbitrary order for special highly degenerate quantum systems.

We will consider $N$-level quantum system with the Hamiltonian $(H_0,V)$:
\begin{equation*}
\label{HN}
H_0=a|1\rangle \langle1|+\sum_{k=2}^N b|k\rangle \langle k|,\quad V=\sum_{k=1}^{N-1}\overline{v}_{k}|k\rangle\langle k+1|+v_{k}|k+1\rangle\langle k|\,.
\end{equation*}
Here $a\neq b$ and  all $v_{k}\in \mathbb{R}$ are nonzero. 
It is known~\cite{SFS,deFouquieresSchirmer} that such a system is completely controllable for any $N$ (a generalization to the case of $v_{k}\in\mathbb C$ for $N=4$ is given in~\cite{KuznetsovPechen}).

\textbf{Theorem.} {\it Let $N\ge 3$, $\rho_0=|N\rangle \langle N|$
and $O=\sum_{k=1}^N \lambda_k |k\rangle\langle k|$ such that $\lambda_1>\lambda_N>\lambda_{N-1}$. Then for any $T\geq T_{\rm min}$ the control $f_0\equiv 0$ is a trap of the $2N-3$ order for $J_O$.}

To prove this statement, note that for such $\rho_0$ and $O$ the complete controllabillity of the quantum system implies that the control $f_0\equiv0$ is not a point of global extremum of $J_O$  for $T\geq T_{\rm min}$~\cite{Pechen2011}. Without loss of generality we can assume that $\lambda_N=0$~\cite{Pechen2011}. 
Let $V_t:=e^{itH_0}Ve^{-itH_0}$. Let $A^n_{lk}\colon \mathfrak{H}^0 \to\mathbb{C}$ be the form of the order $n$ defined as
$$
A^n_{lk}\langle f\rangle:=  \int_0^Tdt_1\int_0^{t_1}dt_2\ldots \int_0^{t_{n-1}}dt_nf(t_1)\ldots f(t_n)\langle l|V_{t_1}\ldots V_{t_n}|k\rangle.
$$
Let $A^0_{lk}=\delta_{lk}$ (the Kronecker symbol). By direct calculations, one can obtain a formula for Fr\'echet differential of order $n$ of the objective functional $J_O$ at~$f_0$:
\begin{equation}
\label{Jm1}
\frac 1{n!}J^{(n)}_O(f_0)(f,\ldots,f)=\sum_{j=0}^n \sum_{l=1}^{N-1} (-1)^{n-j}i^{n} \lambda_l A^j_{lN}\langle f \rangle \overline{A^{n-j}_{lN}\langle f \rangle}
\end{equation}
For $n=1$ we get that $J'_O(f_0)=0$. Due to $\langle l|V|N\rangle=0$  for $l\neq N-1$, the second differential of $J_O$ at $f_0$ has the form $$\frac 1{2!}J''_O(f_0)(f,f)=\lambda_{N-1}|A^1_{(N-1)N}\langle f\rangle|^2=\lambda_{N-1}v^2_{N-1}\left(\int_0^Tf(t)dt\right)^2.$$
Introduce the space
$\mathfrak{H}^{1}=\{f \in \mathfrak{H}^0\colon \int_0^Tf(t)dt=0\}$.
 Then $J''_O(f_0)(f,f)<0$ for $f\in  \mathfrak{H}^0 \setminus\mathfrak{H}^1$ and $J''_O(f_0)(f,f)=0$ for $f\in \mathfrak{H}^1$. Note that the following holds true for the quantum system $(H_0,V)$.
If $1<l$ and $n\leq N-1$, then $\langle l|V_{t_n}\ldots V_{t_1}|N\rangle=\langle l|V^n|N\rangle$  and, hence, the form
$A^n_{lN}\langle f\rangle=\frac{\langle l|V^n|N\rangle}{n!}\left(\int\limits_0^Tf(t)dt\right)^{n}$ vanishes on $\mathfrak{H}^{1}$.
Also, if $n<N-1$, then $\langle 1|V_{t_n}\ldots V_{t_1}|N\rangle=0$, and so $A^n_{1N}=0$. Then it follows from  formula~(\ref{Jm1}) that 
$J^{(n)}_O(f_0)(f,\ldots,f)=0$ for $3\leq n \leq 2N-3$
 and for $f\in \mathfrak{H}^1$. Moreover,
 \begin{multline}
\frac 1{(2N-2)!} J_O^{(2N-2)}(f_0)(f,\ldots,f)=\lambda_1|A_{1N}^{N-1}\langle f \rangle|^2=\\=\lambda_1\Bigl|\int\limits_{[0,T]^{N-1}}K(t_1,\ldots,t_{N-1})f(t_1)\ldots f(t_{N-1})dt_1\ldots dt_{N-1}\Bigr|^2\geq 0
\end{multline}
for $f\in\mathfrak{H}^1$,
where
$K(t_1,t_2,\ldots,t_{N-1})=\frac 1{(N-1)!}v_{1}v_{2}\ldots v_{N-1}e^{i(a-b)\max(t_1,\ldots,t_{N-1})}$.
Thus the  control $f_0\equiv 0$ is a trap of the $2N-3$ order. This completes the proof of the Theorem.

The authors are grateful to S.A.~Kuznetsov for pointing out the proof of controllability in~\cite{SFS}.


\begin{thebibliography}{99}


\bibitem{Koch_2022}
C.P. Koch, U. Boscain, T. Calarco, ~G. Dirr, S. Filipp,
S.J. Glaser, R. Kosloff, S. Montangero, T. Schulte-Herbr\"{u}ggen,
D. Sugny, F.K. Wilhelm, Quantum optimal control in quantum technologies. 
Strategic report on current status, visions and goals for research 
in Europe, {\it EPJ Quantum Technology}, {\bf 9}, 19 (2022).
\url{https://doi.org/10.1140/epjqt/s40507-022-00138-x}

\bibitem{Rabitz2004}
H.A.~Rabitz, M.M.~Hsieh, C.M.~Rosenthal, Quantum optimally controlled transition landscapes, {\it Science}, {\bf 303}, 1998--2001 (2004).
\url{https://doi.org/10.1126/science.1093649}

\bibitem{PechenPRA2012} A. Pechen, N. Il'in, 
Trap-free manipulation in the Landau-Zener system, 
{\it Phys. Rev. A}, {\bf 86}, 052117 (2012).
\url{https://doi.org/10.1103/PhysRevA.86.052117}

\bibitem{PechenRMS2015}  A.N.~Pechen, N.B.~Il'in, 
On critical points of the objective functional for maximization of qubit observables, 
{\it Russian Math. Surveys}, {\bf 70}, 782--784 (2015).
\url{https://doi.org/10.1070/RM2015v070n04ABEH004962}

\bibitem{VolkovJPA2021} B.O.~Volkov, O.V.~Morzhin, A.N.~Pechen,  Quantum control landscape for ultrafast generation of single-qubit phase shift quantum gates,
{\it J. Phys. A},  {\bf 54}, 215303 (2021).
\url{https://doi.org/10.1088/1751-8121/abf45d}

\bibitem{Pechen2011} 
A.N.~Pechen, D.J.~Tannor, Are there traps in quantum control landscapes? {\it Phys. Rev. Lett.}, {\bf 106}, 120402 (2011).
\url{https://doi.org/10.1103/PhysRevLett.106.120402}

\bibitem{Pechen2012} 
A.N.~Pechen, D.J.~Tannor, 
Quantum control landscape for a Lambda-atom in the vicinity of second-order traps, 
{\it Isr. J. Chem.}, {\bf 52}, 467--472 (2012).
\url{https://doi.org/10.1002/ijch.201100165}

\bibitem{deFouquieresSchirmer}
P.~de Fouquieres, S.G.~Schirmer, A closer look at quantum control landscapes and their implication for control optimization, {\it Infin. Dimens. Anal. Quant. Probab. Rel. Top.}, {\bf 16}, 1350021 (2013).
\url{https://doi.org/10.1142/S0219025713500215}

\bibitem{SFS} S.G.~Schirmer, H.Fu, A.I.~Solomon, 
Complete controllability of quantum systems, 
{\it Phys.~Rev.~A}, \textbf{63}, 063410 (2001). 
\url{https://doi.org/10.1088/0305-4470/34/8/313}

\bibitem{KuznetsovPechen} 
S.A.~Kuznetsov,  A.N.~Pechen, On controllability of a highly degenerate four-level quantum system with a ``chained'' coupling Hamiltonian, {\it Lobachevskii J. Math.}, \textbf{43}, 1683-1692 (2022).
\url{https://doi.org/10.1134/S1995080222100225}

\end{thebibliography}
\end{document}